\documentclass[pra,preprint,a4paper,showpacs,superscriptaddress]{revtex4}
\usepackage{amsmath}
\usepackage{amsfonts}
\usepackage{graphicx}
\begin{document} 
\title{Separability bounds on multiqubit moments due to positivity under partial transpose} 
\author{A. R. Usha Devi}
\email{arutth@rediffmail.com}
\affiliation{Department of Physics, Bangalore University, 
Bangalore-560 056, India}
\affiliation{H. H. Wills Physics Laboratory, University of Bristol, Bristol BS8 1TL, UK}
\affiliation{Inspire Institute Inc., McLean, VA 22101, USA.}
\author{A. K. Rajagopal}
\affiliation{Inspire Institute Inc., McLean, VA 22101, USA.} 

\date{\today}

\begin{abstract}
Positivity of the density operator reflects itself in terms of  sequences of 
inequalities on  observable moments. Uncertainty relations for non-commuting observables 
form a subset of these inequalities. In addition, criterion of positivity under partial transposition~(PPT) 
 imposes distinct bounds on moments, violations of which  signal entanglement. 
We present bounds on some novel sets of  composite moments, consequent to  
positive partial transposition of the density operator  and report their violation by entangled 
multiqubit states. In particular, we derive separability bounds on  a multiqubit moment matrix (based on PPT 
constraints on  bipartite divisions of the density matrix) and show that  three qubit pure states with non-zero 
tangle violate these PPT moment constraints. Further, we recover necessary and sufficient condition of 
separability in a multiqubit Werner state through PPT bounds on moments. 
\end{abstract}

\maketitle
\section{Introduction}
Quantum description of nature departs from classical one at various levels.  
Uncertainty relations (UR) exhibit quantum feature in a fundamental manner. Entanglement is another peculiar 
quantum signature, which has evoked much interest from the point of view of its applicability in quantum 
information science~\cite{1} - in addition to promoting a deeper conceptual understanding. While UR bring out 
explicit constraints placed on  first and second moments of two non-commuting observables, it is a complex issue 
to specify how inseparability manifests itself through observable moments. Exploring such practical entanglement 
tests forms an active field of research~\cite{Guhne,SV,Practical,U,Gillet}. 

A hierarchy of inequalities for the observable moments could be formulated  based on the non-negativity of 
the density operator characterizing the quantum system~\cite{Oc, Nar, NM} - uncertainty principle occuring as a 
particular attribute of this property.  In the case of composite quantum systems, appropriately chosen 
matrix of moments associated with the set of all states, which are positive under partial 
transpose~(PPT)~\cite{Peres}, get additional restrictions - thus paving way for distinguishing them from the 
class of quantum entangled states that are non-positive under partial transpose~(PT). 
It may be emphasized  that  the PPT constraints on composite moments are {\em equivalent} to the positivity of 
the associated partially transposed  density matrix (when PT on  bipartite divisions of the system are 
considered).  The reformulation of the problem  of  testing 
entanglement in terms of observable moments has been successful in the case of two-mode gaussian 
states, where necessary and sufficient conditions of entanglement get encrypted as inequalities~\cite{Simon,CV1} 
on second order moments of canonical observables  $(\hat q_1,\hat p_1; \hat q_2,\hat p_2)$. 
It is  a complex issue to find a general prescription to identify appropriate  
basic operators associated with composite systems, which reveal entanglement 
in terms of  optimal bounds on observable moments. However, operators exhibiting 
simple transformation properties under PT  serve as useful tools in exploring how inseparability gets encoded in 
observable moments~\cite{Gillet,Simon,CV1,CV2}.  

The purpose of the present paper is to investigate  separability bounds on composite moments associated with 
multiqubit systems,  by exploiting simple transformation properties of the basic qubit operators under PT.  
 With suitable choices of operators, we recover 
necessary and sufficient conditions  for separability in multiqubit systems of interest - 
formulated here as PPT inequalities on moments. 
More specifically, we recover inseparability conditions of Horodecki et. al.~\cite{Hor} on the 
state parameters of entangled two qubit systems as  violation of  
PPT bounds on a given moment matrix. We construct a generalized multiqubit moment matrix 
 involving basic observables of the system, and show that PPT moment inequalities 
are violated by GHZ-like  pure states. Inseparable multiqubit Werner states 
are also shown to necessarily violate the PPT moment constraints.       

The organization of this paper is as follows: 
In Sec.~II a general description to find PPT  bounds on the 
matrix of moments is outlined. In Sec.~III suitable matrices of two qubit moments 
are investigated and PT restrictions on  moments are derived. Entangled two qubit states 
are shown to violate these PPT bounds on moments.   
In Sec.~IV  a special  $4\times 4$ multiqubit moment matrix is proposed and restriction on it due 
to partial transpose on  bipartite divisions of the system is identified. Entangled three qubit pure states with 
non-zero tangle are shown to necessarily violate the 
PPT bounds on the moments. Further, we recover necessary and sufficient condition for separability 
in  multiqubit Werner state through PPT moment restrictions. Sec. V has a summary of results.     

\section{General description of PPT inequalities on moments}        

Consider a set of linearly independent, hermitian observables 
$\{\hat A_{i},\  i~=~0,1,2,\ldots, \}$, with $\hat A_0=\hat I$ 
being the identity operator,   arranged in the form of a operator column (row)  $\hat \xi$ 
($\hat \xi^T$) as,    
\begin{equation} 
\label{xi}
\hat\xi^T=\left(\begin{array}{cccc} \hat A_0 & \hat A_1 & \ldots & \hat A_n \end{array}\right).
\end{equation} 
The moment matrix~\cite{NM}, 
$M(\hat\rho)={\rm Tr}[\hat\rho\,\hat\xi\, \hat\xi^\dag],$ 
formed by taking quantum averages of operator  entries of $\hat\xi\, 
\hat\xi^\dag$ in a density operator $\hat\rho$ is given by,  
\begin{equation}
\label{MM}
M(\hat\rho)=
\left(\begin{array}{ccccc} 1 & \langle\hat A_1\rangle & 
\langle \hat A_2\rangle & \ldots &  \ldots \cr 
 \langle\hat A_1\rangle &   \langle\hat A_1^2\rangle & \langle\hat A_1\hat A_2\rangle & \ldots  & \ldots \cr 
 \langle\hat A_2\rangle &   \langle\hat A_2\hat A_1\rangle & \langle\hat A_2^2\rangle & \ldots &  \ldots \cr 
 \vdots &  \vdots & \vdots & \vdots & \vdots \cr 
 \vdots &   \vdots & \vdots & 
  \vdots & \vdots   
  \end{array}\right).
  \end{equation}
By virtue of its construction, $M(\hat\rho)\geq 0,$ a condition imposed due to  
the non-negativity of $\hat\rho$. In other words, positive semi-definiteness of the density matrix $\hat\rho$ is 
reformulated in terms of observable moments of all orders. 
The well-known Schrodinger-Robertson~(SR) uncertainty 
relation for the observables $A_1,\ A_2$ i.e., 
\begin{equation}
\label{SR}
\langle(\bigtriangleup \hat A_1)^2\rangle\, \langle(\bigtriangleup \hat A_2)^2\rangle \geq 
\frac{1}{4}\left(\vert \langle [\hat A_1,\hat A_2]\rangle\vert^2+
\langle \{\bigtriangleup\hat 
A_1,\bigtriangleup\hat A_2\}\rangle^2\right),
\end{equation}
(where $\bigtriangleup \hat A_i=\hat A_i-\langle \hat A_i\rangle,$ \ and   
$\{\bigtriangleup\hat 
A_1,\bigtriangleup\hat A_2\}=\bigtriangleup~\hat A_1\bigtriangleup~\hat A_2+ 
\bigtriangleup~\hat A_2\, \bigtriangleup~\hat A_1$) emerges~\cite{NM} as a consequence of the    
positive semi-definiteness of the  $3\times 3$ principle diagonal block of (\ref{MM}):   
$$\left(\begin{array}{ccc} 1 & \langle\hat A_1\rangle & 
\langle \hat A_2\rangle  \cr 
 \langle\hat A_1\rangle &   \langle\hat A_1^2\rangle & \langle\hat A_1\hat A_2\rangle \cr 
 \langle\hat A_2\rangle &   \langle\hat A_2\hat A_1\rangle & \langle\hat A_2^2\rangle 
 \end{array}\right)\geq 0.$$ 
For the canonical observables $\hat A_1=\hat q,\ \hat A_1=\hat p$, 
satisfying the commutation relation $[\hat q,\hat p]=i\, \hbar,$ this leads to an unsurpassable quantum limit  
$$\langle(\bigtriangleup \hat q)^2\rangle\, \langle(\bigtriangleup \hat p)^2\rangle-
\frac{1}{4}\, \langle \{\bigtriangleup\hat q,\bigtriangleup\hat p \}\rangle^2 \geq \frac{\hbar^2}{4},$$ 
which serves as both necessary and sufficient condition for a single mode  Gaussian to be a legitimate  
quantum state. When more general states are considered it becomes a nontrivial task to 
identify a finite set of  inequalities on moments,  capturing the positivity of 
the quantum states {\em completely}.     

If a given multipartite quantum state is   
separable, i.e., when the state can be expressed as a convex sum of product states,    
\begin{equation*}
\hat\rho^{\rm sep}=\sum_i\, p_i\, \hat\rho_{1i}\otimes \hat\rho_{2i}\otimes \ldots  \hat\rho_{ni},\ \ \ 
\sum_i p_i=1,\ \ 0\leq p_i\leq 1,  
\end{equation*}
PT operation~\cite{Peres} on any bipartite division of the state preserves its hermiticity, positive 
semi-definiteness and unit trace, leading to another physically valid separable state. 
The moment matrix $M(\hat\rho^{\rm sep})$ as well as the corresponding matrix 
$M((\hat\rho^{\rm sep})^{\rm PT}),$ constructed using partially 
transposed separable density operator are bound to be non-negative. 
Therefore, separability implies additional restriction viz., $M(\hat\rho^{PT})\geq 0$
on the moments, which is stronger, in general, than the usual moment 
matrix positivity condition $M(\hat\rho)\geq 0$. 

If the PT map is transferred to operators   i.e.,   
\begin{equation}
{\rm Tr}[\hat\rho^{\rm PT}{\cal O}]={\rm Tr}[\hat\rho\, {\cal O}^{\rm PT}]\ \ \ \  
{\rm with}\ \ {\cal O}\longrightarrow {\cal O}^{\rm PT},
\end{equation}
positivity of the moment matrix  in a separable state may be expressed as, 
\begin{eqnarray}
M((\hat\rho^{\rm sep}))\geq 0, \ \  M^{\rm PT}(\hat\rho^{\rm sep})\geq 0,   
\end{eqnarray}  
where $ M^{\rm PT}(\hat\rho^{\rm sep})=M((\hat\rho^{\rm sep})^{\rm PT})$.  

An equivalent of SR inequality (\ref{SR}) in a partially transposed separable state 
is readily obtained by demanding positivity of the $3\times 3$ principle diagonal block of  
$M^{\rm PT}(\hat\rho^{\rm sep})$:
\begin{widetext}
\begin{equation}
\label{SRPT}
\langle((\bigtriangleup \hat A_1)^2)^{\rm PT}\rangle\, \langle((\bigtriangleup \hat 
A_2)^2)^{\rm PT}\rangle \geq \frac{1}{4}\, \vert \langle [\hat A_1,\hat A_2]^{\rm 
PT}\rangle\vert^2+\frac{1}{4}\, \langle \{\bigtriangleup\hat 
A_1,\bigtriangleup\hat A_2\}^{PT}\rangle^2 
\end{equation}
\end{widetext}
Gillet et. al.~\cite{Gillet} considered special observables  $\hat A_1$ and $\hat A_2$ 
satisfying the property 
\begin{eqnarray}
\label{sp}
(\hat A_1^2 )^{\rm PT}=(\hat A_1^{\rm PT})^2,\  (\hat A_2^2 )^{\rm PT}=(\hat A_2^{\rm PT})^2,
\end{eqnarray}
in which case the LHS of  the inequality 
(\ref{SRPT}) involves a product of variances $\langle(\bigtriangleup\hat A_1^{\rm PT})^2\rangle\, 
\langle(\bigtriangleup\hat A_1^{\rm PT})^2\rangle$ and the resulting Schrodinger-Robertson partial transpose 
(SRPT) inequality   is, in general, stronger compared to the  
traditional SR uncertainty for the operators $A_1^{\rm PT},\ A_2^{\rm PT}$. 
SRPT inequality~\cite{Gillet} is necessarily obeyed 
by the set of all separable states and its  violation   
would therefore be sufficient to detect entanglement. Using  special observables satisfying (\ref{sp}), 
  a wide class of entangled pure bipartite, tripartite states 
of qubits, angular momentum states of harmonic oscillators, cat states, etc. are shown~\cite{Gillet} 
to violate the SRPT inequality.  In general, violation of PPT bounds  $M^{\rm PT}((\hat\rho^{\rm sep}))~\geq~ 0$ 
provide a series of constraints on observable moments, which leads to operational tests of 
 entanglement.

It is worth noting that the implications of PT on observables is established as  
"local time-reversal" in the case of canonical pair of observables 
$\{\hat q_\alpha,\ \hat p_\alpha\}$ of continuous variable~(CV) states and also for basic  qubit observables
$\vec\sigma_\alpha$: 
\begin{eqnarray}
\label{ptmapcv}
{\rm PT\  on\ CV\ observables\ } &{\rm (w.r.t}& \  \alpha^{\rm th}{\rm \ subsystem):} \ \ \nonumber \\
\hat q_\alpha,\ \hat p_\alpha & \longrightarrow&  \hat q_\alpha,\ -\hat p_\alpha \nonumber  \\
\hat q_\beta,\ \hat p_\beta  &\longrightarrow&  \hat q_\beta,\ \hat p_\beta,\ \ \ \beta\neq \alpha   \\
\ {\rm PT\ on\ qubit\ observables:} &\hskip 0.1in& \nonumber \\
\label{ptmapqu}
\vec\sigma_\alpha &\longrightarrow &  -\vec\sigma_\alpha \nonumber \\
\vec\sigma_\beta  & \longrightarrow & \vec\sigma_\beta,\ \ \ \beta\neq \alpha. 
\end{eqnarray}  

For bipartite CV states, non-negativity of the  moment matrix 
$M(\hat\rho)~=~\langle\hat\xi\hat\xi^\dag\rangle$  constructed using the 
operator column $\hat\xi^T~=~(\hat I,\hat \varsigma^T)$ of basic canonical pair of observables 
$\hat\varsigma^T=(\hat q_1,\hat p_1; \hat q_2,\hat p_2)$, leads to  uncertainty condition 
on the variance matrix of the two mode CV state: 
\begin{equation}
\label{vunc}
V+\frac{i}{2}\, \beta\geq 0,
\end{equation} 
where $V$ denotes the $4\times 4$ real symmetric variance matrix~\cite{SM} with elements,  
$V_{ab}=\frac{1}{2}\langle\{\bigtriangleup\hat\varsigma_a,\bigtriangleup\hat\varsigma_b\}\rangle$  
and the matrix  $\beta$ is defined through  $i\,(\beta)_{ab}=\langle[\hat\varsigma_a,\hat \varsigma_b]\rangle.$ 
The PT map on the canonical  observables,  
leads to a  structurally similar uncertainty-like restriction   
on the variance matrix expressed compactly as~\cite{Simon}
\begin{equation}
\label{vptunc}
V^{\rm PT}+\frac{i}{2}\beta\geq 0, \ \ \ V^{\rm PT}=\Omega V \Omega,
\end{equation}  
where $\Omega={\rm diag}\, (1,-1,1,1),$ for partial transpose taken on the first system.   
Separable bipartite CV states never violate the inequality (\ref{vptunc}) and  its 
violation signals entanglement of a bipartite CV state. Moreover, violation of 
(\ref{vptunc}) serves as both necessary and sufficient inseparability condition for an arbitrary two mode 
Gaussian state~\cite{Simon}. 
Inequalities on higher order moments of bipartite CV states, involving canonical pairs of observables 
 have been formulated by Shchukin and Vogel~(SV)~\cite{SV}, based on   
the positivity of corresponding moment matrix $M^{\rm PT}(\hat\rho)$. SV result provided 
a common basis for many other CV inseprability criteria~\cite{Simon, CV1, CV2}  
derived previously, including those, which  appeared to be independent of the PPT condition. 

In the following, we investigate the PPT bounds on some specially constructed  
 moment matrices associated with multiqubit operators, which exhibit well defined PT 
maps and explore violation of  PPT bounds imposed on them.

\section{Two qubit PPT moment inequalities}   
An arbitrary two-qubit density operator belonging to  
the Hilbert-Schmidt space ${\cal H}={\cal C}^2\otimes{\cal C}^2$ has the form, 
\begin{equation}
\rho=\frac{1}{4}\, [I\otimes I+ \vec\sigma\cdot\vec{s}^{(1)}\otimes I + 
I\otimes \vec\sigma\cdot\vec{s}^{(2)}+ 
          \sum_{i,j=1,2,3} \sigma_{1i}\otimes \sigma_{2j}\, t_{ij}\,]\, , 
\label{rho} 
\end{equation}
where      $\sigma_i$ are the standard Pauli spin matrices; $I$ denotes the $2\times 2$ unit matrix.  
Here, qubit averages are given by,  
\begin{equation}
\label{s}
\vec{s}^{(1)}={\rm Tr}\,(\rho\,  \vec\sigma\otimes I), \ \vec{s}^{(2)}={\rm Tr}\,(\rho\,  I\otimes\vec\sigma)
\end{equation} 
and
\begin{equation}
\label{t}  
t_{ij}={\rm Tr}\,(\rho\, \sigma_{i}\otimes \sigma_{j}),
\end{equation} 
are  elements of the real $3\times 3$ two qubit correlation matrix   
\begin{equation}
\label{Tmatrix}
T=\left(\begin{array}{ccc} t_{11} & t_{12} & t_{13}\cr 
t_{21} & t_{22} & t_{23} \cr 
t_{31} & t_{32} & t_{33} \end{array}\right).  
\end{equation} 

A direct examination of positivity bounds on the moment matrix may be done   
by  considering a  basic column of qubit operators   
$\hat\xi^T_1 ~=~(\begin{array}{ccc} I\otimes I, & \sigma_i\otimes I,&   I\otimes \sigma_j \end{array}).$ We 
construct 
a real symmetric  moment matrix $M_1(\hat\rho)~=~\langle \hat\xi_1\hat\xi_1^\dag\rangle$, expressed in a 
suitable block form as,
\begin{eqnarray}
M_1(\hat\rho)&=&\left(\begin{array}{ccc} 1 & s^{(1)T} & s^{(2)T} \cr s^{(1)} & {\cal I} & T \cr s^{(2)} & T^T & 
{\cal I} \end{array}\right) =\left(\begin{array}{cc} 1 & C^T \cr C & B\end{array}\right), \nonumber \\ 
{\rm where}\ \  B&=& \left(\begin{array}{cc} {\cal I} & T \cr T^T & {\cal I}\end{array}\right),\ \ 
 C^T= \left(\begin{array}{cc} s^{(1)T}, & s^{(2)T}\end{array}\right)
\end{eqnarray} 
where $s^{(1)},\ s^{(2)}$ denote three componental columns of qubit averages: $s^{(1)T}~=~(\begin{array}{ccc} 
s^{(1)}_{1},& s^{(1)}_2, & s^{(1)}_3\end{array}), \ s^{(2)T}~=~(\begin{array}{ccc} s^{(2)}_{1},& s^{(2)}_2, & 
s^{(2)}_3\end{array}), $ and ${\cal I}$ denotes $3\times 3$ identity matrix.  

A congruence transformation 
\begin{equation}
M_1(\hat\rho)\longrightarrow L\, M_1(\hat\rho)\, L^T =
\left(\begin{array}{cc}   1 & 0 \cr  0 & B-C\, C^T \end{array}\right)  
\end{equation}
with
$L~=~\left(\begin{array}{cc} 1 & 0 \cr -C & {\cal I}\oplus {\cal I} \end{array}\right)$ 
leads to the identification that   
\begin{eqnarray}
\label{momcon}
&&M_1(\hat\rho)\geq 0 \Longleftrightarrow B-C\,C^T \geq 0\hskip 1.5in\nonumber \\  
&&\ {\rm i.e.,}\  \left(\begin{array}{cc}  
{\cal I}- s^{(1)} s^{(1)T}& T-s^{(1)} s^{(2)T} \cr  T^T-s^{(2)} s^{(1)T} & {\cal I}- s^{(2)} s^{(2)T} 
\end{array}\right)\geq 0.
\end{eqnarray}
The corresponding PT moment matrix $M^{\rm PT}_1(\hat\rho)\equiv M_1(\hat\rho^{\rm PT}) $   is 
obtained by using (\ref{ptmapqu}) (with PT operation on the first qubit) and after some simple algebra we 
obtain, 
\begin{equation}
\label{2qmpt}
M^{PT}_1(\hat\rho)=\left(\begin{array}{ccc} 1 & -s^{(1)T} & s^{(2)T} \cr -s^{T} & {\cal I} & -T 
\cr s^{2} & -T^T & {\cal I} \end{array}\right)\ 
 \end{equation} 
Following similar arguments illustrated above, we identify that
\begin{eqnarray}
\label{momcon2}
M^{\rm PT}_1(\hat\rho)\geq 0 \Longleftrightarrow \hskip 1.8in \nonumber \\ 
\nonumber \\
\left(\begin{array}{cc}  
{\cal I}- s^{(1)} s^{(1)T}& -(T-s^{(1)} s^{(2)T}) \cr  -(T^T-s^{(2)} s^{(1)T}) & {\cal I}- s^{(2)} s^{(2)T} 
\end{array}\right)\geq 0.
\end{eqnarray}
An examination of (\ref{momcon}) and (\ref{momcon2}) reveals that for  two qubit states with 
random subsystems i.e., $s^{(1)}=0=s^{(2)},$ we have 
  $M_1(\hat\rho)~\geq~ 0 \Rightarrow M^{PT}_1(\rho)\geq 0.$  In  other words, the PPT bound (\ref{momcon2}) does 
not impose stronger restriction on moments other than the usual positivity constraints (\ref{momcon})  and 
thus, it fails to capture the inseparability  of  the two qubit state with  disordered subsystems.

A suitable moment matrix,  PPT constraints on which allow a clear distinction between separable and entangled 
two qubit states, could indeed be realized and is discussed in the following: 
 We construct  a $4\times 4$ moment matrix $M_2(\hat\rho)={\rm Tr}[\hat\rho\,\hat\xi_2\, 
\hat\xi_2^\dag],$ 
where the operator column $\hat\xi_2$ is chosen to be, 
\begin{equation}
\label{xi2}
\hat\xi_2= \left(\begin{array}{c} I\otimes I \cr
\vec \sigma\cdot \vec{k}_1\otimes \vec \sigma\cdot \vec{k}_2 \cr 
\vec \sigma\cdot \vec{l}_1\otimes\vec \sigma\cdot \vec{l}_2\cr
\vec \sigma\cdot \vec{m}_1\otimes\vec \sigma\cdot \vec{m}_2
 \end{array}\right)   
\end{equation} 
with $\{\vec{k}_1,\vec{l}_1,\vec{m}_1\}$ and $\{\vec{k}_2,\vec{l}_2,\vec{m}_2\}$   
denoting two sets of mutually orthogonal real three dimensional unit vectors. 
 The moment matrix associated with the operator column (\ref{xi2}) takes the following explicit form: 
\begin{equation}
M_2(\hat\rho)=\left( \begin{array}{cccc} 1 & t_{k_1k_2} & t_{l_1l_2} & t_{m_1m_2} \cr 
t_{k_1k_2} &  1 & -t_{m_1m_2} & -t_{l_1l_2} \cr 
t_{l_2l_2} & -t_{m_1m_2} & 1 & -t_{k_1k_2} \cr 
t_{m_1m_2} & -t_{l_1l_2} & -t_{k_1k_2} & 1 \end{array}\right),
\end{equation} 
where we have denoted $t_{k_1 k_2}=k_1^TTk_2$, $ t_{l_1 l_2}~=~l_1^TTl_2$, 
$t_{m_1 m_2}=m_1^TTm_2$  with $T$ corresponding to  the two qubit correlation matrix given by 
~(\ref{Tmatrix}); $k_\alpha,l_\alpha,m_\alpha,\ \alpha=1,2$ being 3 componental columns 
involving components of unit vectors $\vec{k}_\alpha,\, \vec{l}_\alpha, \vec{m}_\alpha$.   

Partial transpose operation with respect to first qubit leads to the partial {\em time reversal} transformation 
(\ref{ptmapqu}) on the qubit operators and the corresponding moment matrix is given explicitly by,  
\begin{equation}
\label{PPTB}
M^{\rm PT}_2(\hat\rho)=\left( \begin{array}{cccc} 1 & -t_{k_1k_2} & -t_{l_1l_2} & -t_{m_1m_2} \cr 
-t_{k_1k_2} & 1 &   t_{m_1m_2} & t_{l_1l_2}\cr 
-t_{l_1l_2} & t_{m_1m_2} & 1 & t_{k_1k_2} \cr 
-t_{m_1m_2} & t_{l_1l_2} & t_{k_1k_2} & 1 \end{array}\right).
\end{equation} 
The PPT bound  $M^{PT}_2(\hat\rho)\geq 0$ is violated if any of the eigenvalues      
\begin{eqnarray}
\label{pptb2}
\mu_1^{\rm PT}&=&1 + t_{k_1k_2} -t_{l_1l_2} - t_{m_1m_2}\nonumber \\ 
 \mu_2^{\rm PT}&=& 1 - t_{k_1k_2} + t_{l_1l_2} - t_{m_1m_2}\nonumber \\ 
 \mu_3^{\rm PT}&=&  1 - t_{k_1k_2} - t_{l_1l_2} + t_{m_1m_2} \\ 
 \mu_4^{\rm PT}&=&   1 + t_{k_1k_2} + t_{l_1l_2} + t_{m_1m_2}.\nonumber
\end{eqnarray} 
of  (\ref{PPTB}) assume negative values. We illustrate the power of this choice by way of examples.   

Consider a two-qubit Werner state 
\begin{equation}
\hat\rho_{W}=\frac{(1-x)}{4}\, I\otimes I +x\, \vert \Psi^{(-)}\rangle\langle  \Psi^{(-)}\vert,\ 0\leq x\leq 1 
\end{equation}  
where $\vert \Psi^{(-)}\rangle=\frac{1}{\sqrt{2}}[\vert 0,1\rangle-\vert 1,0\rangle]$ is a two qubit Bell state.  
The two qubit correlation matrix is readily found to be  
$T={\rm diag}(-x,-x, -x)$ and with a choice 
$k_1^T=\left(\begin{array}{ccc}1, & 0, & 0\end{array}\right)=k_2^T,$ 
 $l_1^T~=~\left(\begin{array}{ccc}0,& 1, & 0\end{array}\right)~=~l_2^T,$
\ $m_1^T~=~\left(\begin{array}{ccc}0, & 0, & 1\end{array}\right)=m_2^T,$
 
it is easy to find the eigenvalues of   $M^{\rm PT}_2(\rho_W):$ 
\begin{equation}
\mu_1^{PT}=1-3x,\ \mu_2^{\rm PT}=\mu_3^{\rm PT}=\mu_4^{\rm PT}=1+x.
\end{equation}  
Clearly, $M^{\rm PT}_2(\rho_W)$ respects the PPT bound, 
when $0~\leq~ x~\leq~ \frac{1}{3},$ which is the well-known 
separability domain~\cite{Peres} for the  two qubit Werner state.

We also recover the inseparability conditions  
of Horodecki et. al.,~\cite{Hor}  
through  non-positivity of the eigenvalues (\ref{pptb2}). In order to see this, 
we recall that the two qubit state parameters 
$\{  s^{(1)}_i,\ s^{(2)}_i,   \  t_{ij}\}$ (defined in   
(\ref{s}),(\ref{t})) transform under local unitary 
operations $U_1\otimes U_2$ on the qubits as follows~\cite{Hor}: 
\begin{eqnarray}
\label{tran} 
s^{(1)'}_i=\displaystyle\sum_{j=x,y,z} O^{(1)}_{ij}\,  s_{1j},   &  s'_{2i}=
\displaystyle\sum_{j=x,y,z} O^{(2)}_{ij}\, s_{2j} \,,\nonumber\\
 t'_{ij}=\sum_{k,l=x,y,z} O^{(1)}_{ik}\, O_{jl}^{(2)}t_{kl} &  {\rm \ or \ }  
   \,\, T'=O^{(1)}\, T\,  O^{(2)\, T} ,  
    \end{eqnarray}    
where $O^{(\alpha)}\in SO(3)$ denote the  $3\times 3$ real orthogonal rotation matrices, 
corresponding uniquely to the $2\times 2$ unitary matrices $U_{\alpha}~\in~SU(2).$   
As the entanglement properties of the two qubit state remain unaltered under local unitary operations 
one may choose to specify the state parameters $\vec s^{(\alpha)},\ (\alpha=1,2)$ and $T$ in a basis in which
$T$ is diagonal. This is possible because
the real $3\times 3$ correlation matrix $T$ can always be transformed to a diagonal form with the help 
of an appropriate local transformation $U_1\otimes U_2$:
\begin{equation}
\label{diagT}
T^d=O^{(1)}\, T\, O^{(2)\, T}={\rm diag}\, (t_1,\, t_2,\, t_3).
\end{equation}

Let us arrange the mutually orthogonal unit column vectors 
$k_\alpha,l_\alpha,m_\alpha, \alpha=1,2$ (involved in the definition (\ref{xi2}) of the operator column 
$\hat\xi_2$) to form $3\times 3$ 
real orthogonal matrices 
\begin{equation}
O^{(1)}=\left(\begin{array}{c} k_1^T \\ l_1^T \\ m_1^T \end{array} \right),\ \ 
O^{(2)}=\left(\begin{array}{c} k_2^T \\ l_2^T \\ m_2^T \end{array} \right).
\end{equation}
With an appropriate choice of unit vectors 
$k_\alpha,l_\alpha,m_\alpha$ we may thus transform 
the two qubit correlation matrix $T$ to its diagonal form  
\begin{eqnarray*}
T^{\rm d}&=&O^{(1)}\,T\,O^{(2)T}\\ 
&=&{\rm diag}\,\left( \begin{array}{ccc} t_{k_1k_2}=t_1, & t_{l_1l_2}=t_2, & t_{m_1m_2}=t_3
\end{array}\right). \\
\end{eqnarray*}  
It is  readily found that the non-positivity of the eigenvalues (\ref{pptb2}) of the PT matrix of moments 
$M^{\rm PT}_2(\hat\rho)$ reduce to the inseparability conditions of Horodecki et. al.~\cite{Hor} i.e.,   
\begin{eqnarray}
\label{H}
1 + t_1 - t_2 - t_3&<& 0,\ \   1 - t_1 + t_2 - t_3< 0, \nonumber \\  
1 - t_1 - t_2 + t_3&<&0, \ \   1 + t_1 + t_2 + t_3< 0.
\end{eqnarray}

\section{PPT bounds on a multiqubit moment matrix}

Now we discuss a special multiqubit moment matrix, which has a distinct and  stronger PPT bound than the 
conventional one. Important class of entangled pure and mixed multiqubit states are shown to violate this PPT 
bound. 
  
Consider the following four componetal column (row) of $N$-qubit operators:  
\begin{equation}
\hat\xi^T=\left(\displaystyle\bigotimes_{\alpha=1}^{N}I,\ \Sigma_1,\ \Sigma_2,\ \Sigma_3\right)
\end{equation} 
where
\begin{eqnarray}
\label{sigma}
 \Sigma_1 &=& \displaystyle\bigotimes_{\alpha=1}^{N}\, \sigma^{k_\alpha 
l_\alpha}_{+}+\displaystyle\bigotimes_{\alpha=1}^{N}\, 
\sigma^{k_\alpha l_\alpha}_{-}
 \nonumber \\
\Sigma_2 &=& -i\,\left(\displaystyle\bigotimes_{\alpha=1}^{N}\, \sigma^{k_\alpha 
l_\alpha}_{+}-\displaystyle\bigotimes_{\alpha=1}^{N}\, \sigma^{k_\alpha l_\alpha}_{-}\right)\nonumber \\
\Sigma_3 &=& \frac{1}{2^N}\, \left(\displaystyle\bigotimes_{\alpha=1}^{N}(I+\vec\sigma\cdot \vec m_\alpha)- 
\displaystyle\bigotimes_{\alpha=1}^{N}(I-\vec\sigma\cdot \vec m_\alpha)\right).\nonumber \\ 
\end{eqnarray}     
Here, we have denoted 
 \begin{equation}
 \label{sigma2}
 \sigma^{k_\alpha l_\alpha}_{\pm}=\frac{1}{2}\, \vec\sigma\cdot 
 [\vec k_\alpha\pm i\, \vec l_\alpha]
 \end{equation}
 and  $ \{\vec{k}_\alpha,\ \vec{l}_\alpha,\ \vec m_\alpha\},\ \alpha=1,2,\ldots, N$ correspond to sets of 
mutually orthogonal three dimensional (real) unit vectors. 

 We give below a more explicit structure of the operators (\ref{sigma}) in the case of two and three qubits:     
\begin{eqnarray*}
{\rm Two\ qubits:}& & \\
 \Sigma_1 &=& \frac{1}{2}\left((\vec\sigma\cdot\vec k_1)\otimes(\vec\sigma\cdot\vec k_2)-
 (\vec\sigma\cdot\vec l_1)\otimes (\vec\sigma\cdot\vec l_2)\right)  \\
\Sigma_2 &=& \frac{1}{2}\left((\vec\sigma\cdot\vec k_1)\otimes(\vec\sigma\cdot\vec l_2)
+(\vec\sigma\cdot\vec l_1)\otimes(\vec\sigma\cdot\vec k_2)\right)  \\
\Sigma_3 &=& \frac{1}{2}\left(\vec\sigma\cdot\vec m_1\otimes I+I\otimes\vec\sigma\cdot\vec m_2\right)  \\
{\rm Three\ qubits:}& & \\
\Sigma_1 &=& \frac{1}{4}\left((\vec\sigma\cdot\vec k_1)\otimes(\vec\sigma\cdot\vec k_2)
\otimes(\vec\sigma\cdot\vec k_3)-(\vec\sigma\cdot\vec l_1)\otimes
(\vec\sigma\cdot\vec l_2)\otimes(\vec\sigma\cdot\vec k_3)\right.\\
&&  -(\vec\sigma\cdot\vec k_1)\otimes(\vec\sigma\cdot\vec l_2)\otimes
(\vec\sigma\cdot\vec l_3)
-(\vec\sigma\cdot\vec l_1)\otimes(\vec\sigma\cdot\vec k_2)\otimes(\vec\sigma\cdot\vec l_3)\\
&& \left. +i\, (\vec\sigma\cdot\vec l_1)\otimes(\vec\sigma\cdot\vec l_2)\otimes
(\vec\sigma\cdot\vec l_3)\right) \\
 \Sigma_2 &=& \frac{1}{4}\left((\vec\sigma\cdot\vec k_1)\otimes(\vec\sigma\cdot\vec l_2)
\otimes(\vec\sigma\cdot\vec k_3)-(\vec\sigma\cdot\vec l_1)\otimes
(\vec\sigma\cdot\vec k_2)\otimes(\vec\sigma\cdot\vec l_3)\right.\\
&&\left. -(\vec\sigma\cdot\vec k_1)\otimes(\vec\sigma\cdot\vec k_2)
\otimes(\vec\sigma\cdot\vec l_3)\right)\\
 \Sigma_3 &=& \frac{1}{4}\left((\vec\sigma\cdot\vec m_1)\otimes I\otimes I +
I \otimes(\vec\sigma\cdot\vec m_2)\otimes I+I\otimes I\otimes(\vec\sigma\cdot\vec m_3)\right. \\
&& \left. +(\vec\sigma\cdot\vec m_1)
\otimes(\vec\sigma\cdot\vec m_2)\otimes(\vec\sigma\cdot\vec m_3)\right).
\end{eqnarray*}

The operators (\ref{sigma}) satisfy the following properties 
\begin{eqnarray}
\Sigma_i\Sigma_j&=& i\, \, \epsilon_{ijk}\, \Sigma_k, \hskip 0.5in  i\neq j=1,2,3 \nonumber  \\
 \Sigma_i^2&=&\frac{1}{2^N}\, \left(\displaystyle\bigotimes_{\alpha=1}^{N}(I+\vec\sigma_\alpha\cdot \vec 
m_\alpha)+ 
\displaystyle\bigotimes_{\alpha=1}^{N}(I-\vec\sigma_\alpha\cdot \vec m_\alpha)\right) \nonumber \\
&&\ \ \ \ i=1,2,3. 
\end{eqnarray}
As will be shown in the foregoing, this novel construction encodes the separability properties 
of  pure three qubit states with non-zero tangle and also in mixed $N$-qubit Werner state.

 Partial transpose on, say first $r$ qubits, corresponds to the following PT map  
on the operators  (\ref{sigma}) (see Eq.~(\ref{ptmapqu})):
 \begin{eqnarray}
\label{sigma3}
\Sigma_1^{\rm PT}&=& (-1)^r\, \Sigma_1,\ \Sigma_2^{\rm PT}= (-1)^r\, \Sigma_2,\nonumber \\ 
\Sigma_3^{\rm PT}&=& \frac{1}{2^N}\, \left(\displaystyle\bigotimes_{\alpha=1}^{r}(I-\vec\sigma_\alpha\cdot 
\vec m_\alpha)\,\displaystyle\bigotimes_{\nu=r}^{N}(I+\vec\sigma_\nu\cdot \vec m_\nu) \right. \nonumber \\
&& \ \ \left. - 
\displaystyle\bigotimes_{\alpha=1}^{r}(I+\vec\sigma_\alpha\cdot \vec 
m_\alpha)\,\displaystyle\bigotimes_{\nu=r}^{N}(I-\vec\sigma_\nu\cdot \vec m_\nu)\right), \nonumber \\
(\Sigma_i\Sigma_j)^{\rm PT} &=&i \, \, \epsilon_{ijk}\, \Sigma_k^{\rm PT},\hskip 0.5in\, i\neq j=1,2,3 
\nonumber \\
(\Sigma_i^2)^{\rm PT}&=&\frac{1}{2^N}\, 
\left(\displaystyle\bigotimes_{\alpha=1}^{r}(I-\vec\sigma_\alpha\cdot 
\vec m_\alpha)\,\displaystyle\bigotimes_{\nu=r}^{N}(I+\vec\sigma_\nu\cdot \vec m_\nu)\right. \nonumber \\ 
&& \ \ \ \left. + 
\displaystyle\bigotimes_{\alpha=1}^{r}(I+\vec\sigma_\alpha\cdot \vec 
m_\alpha)\,\displaystyle\bigotimes_{\nu=r}^{N}(I-\vec\sigma_\nu\cdot \vec m_\nu)\right),\nonumber \\
&  & \ \ i=1,2,3.
\end{eqnarray} 
The moment matrix $M(\hat\rho)$ constructed with the set of operators 
(\ref{sigma}) and its PT analogue $M^{\rm PT}(\hat\rho)$ have identical structures: 
\begin{equation}
M(\hat\rho)=\left(\begin{array}{ccccc}
1 & \langle\Sigma_1\rangle & \langle\Sigma_2\rangle & \langle\Sigma_3\rangle \cr 
\langle\Sigma_1\rangle & \langle\Sigma_0\rangle & i\langle\Sigma_3\rangle 
& -i\langle\Sigma_2\rangle \cr
\langle\Sigma_2\rangle & -i\langle\Sigma_3\rangle & \langle\Sigma_0\rangle & 
i\langle\Sigma_1\rangle \cr
\langle\Sigma_3\rangle & i\,\langle\Sigma_2\rangle & -i\,\langle\Sigma_1\rangle & 
\langle\Sigma_0\rangle \cr 
\end{array}\right),\ \ 
\end{equation}
\begin{equation}
M^{PT}(\hat\rho)=\left(\begin{array}{ccccc}
1 & \langle\Sigma_1^{\rm PT}\rangle & \langle\Sigma_2^{\rm PT}\rangle & \langle\Sigma_3^{PT}\rangle \cr 
\langle\Sigma_1^{\rm PT}\rangle & \langle\Sigma^{\rm PT}_0\rangle & i\langle\Sigma_3^{\rm PT}\rangle & 
-i\langle\Sigma_2^{\rm PT}\rangle \cr
\langle\Sigma_2^{\rm PT}\rangle & -i\,\langle\Sigma_3^{\rm PT}\rangle & \langle\Sigma_0^{\rm PT}\rangle 
& i \langle\Sigma_1^{\rm PT}\rangle \cr
\langle\Sigma_3^{PT}\rangle & i \langle\Sigma_2^{\rm PT}\rangle & -i\langle\Sigma_1^{\rm PT}\rangle & 
\langle\Sigma_0^{\rm PT}\rangle \cr 
\end{array}\right), 
\end{equation}  
where we have denoted  $\Sigma_1^2=\Sigma_2^2=\Sigma_3^2=\Sigma_0.$
 
The eigenvalues of  $M^{\rm PT}(\hat\rho)$ are given  by, 
\begin{widetext}
 \begin{eqnarray}
 \label{eigpt}
\mu_{1\pm}^{\rm PT}&=&\langle\Sigma_0^{\rm PT}\rangle\pm
\sqrt{\langle\Sigma_1\rangle^2+\langle\Sigma_2\rangle^2
+\langle\Sigma_3^{\rm PT}\rangle^2}, \nonumber \\ 
\mu_{2\pm}^{\rm PT}&=& \frac{1}{2}\,(1+\langle\Sigma_0^{\rm PT}\rangle)
 \pm \frac{1}{2}\, \left[(1+\langle\Sigma_0^{\rm 
PT}\rangle)^2+4(\langle\Sigma_1\rangle^2+\langle\Sigma_2\rangle^2
+\langle\Sigma_3^{\rm PT}\rangle^2
-\langle\Sigma_0^{\rm PT}\rangle)\right]^{\frac{1}{2}}
 \end{eqnarray}
 \end{widetext} 
It may be readily seen (from (\ref{eigpt})) that the PPT bound $M^{\rm PT}(\hat\rho)\geq 0$ 
imposes the following restriction on separable states:  
 \begin{equation}
 \label{vio}
\langle\Sigma_0^{\rm PT}\rangle\geq \sqrt{\langle\Sigma_1\rangle^2+\langle\Sigma_2\rangle^2
+\langle\Sigma_3^{\rm PT}\rangle^2}.
\end{equation} 
Entangled multiqubit states 
will be shown to violate the PPT constraint (\ref{vio}) on the moments. 
Specializing (\ref{sigma}) for a random pair 
of qubits drawn from a  symmetric multiqubit system, it is found~\cite{UUR} that spin squeezing~\cite{KU} 
occurs as a consequence of violation of the PPT bound (\ref{vio}). This will be addressed in a separate 
communication~\cite{UUR}.  

We  consider here the example of an arbitrary pure state of three qubits expressed in the 
Schmidt decomposed form as ~\cite{Acin} 
\begin{eqnarray}
\label{3qu}
\vert\Psi\rangle & =&  \lambda_0\vert 0,0,0\rangle+\lambda_1 e^{i\phi}\vert 1,0,0\rangle
\nonumber \\
& &  +\lambda_2\vert 1,0,1\rangle+\lambda_3\vert 
1,1,0\rangle+\lambda_4\vert 1,1,1\rangle
 \end{eqnarray}
 where  $\lambda_i\geq 0, \ \ 0\leq \phi\leq \pi,\  \sum_i\lambda_i^2=1.$
  
Choosing the unit vectors 
\begin{equation}
\label{uv}
\vec k_\alpha=(1,0,0),\ \vec l_\alpha=(0,1,0),\ \vec m_\alpha=(0,0,1) 
\end{equation}
in Eqs.~(\ref{sigma}), (\ref{sigma2}),  
 we  obtain (with a PT operation on the first qubit),  
\begin{eqnarray}
\langle\Sigma_1\rangle &=& \lambda_0\,\lambda_4,\ \langle\Sigma_2\rangle=0,\nonumber \\  
\langle\Sigma_3^{PT}\rangle&=& \lambda_1^2,\ \ \langle(\Sigma_0^{\rm PT})\rangle= \lambda_1^2. 
\end{eqnarray} 
The PPT inequality (\ref{vio}) is violated if and only if   
\begin{equation}
\label{tan}
\lambda_0\,\lambda_4 > 0.
\end{equation}
Recalling that the 3-tangle~\cite{Coffman}, a measure of genuine three qubit entanglement, is given by
\begin{equation} 
\tau_3=4\lambda_0^2\lambda_4^2
\end{equation}  
we find an interesting result: {\em  the PPT bound (\ref{vio}) is violated iff an arbitrary three qubit state 
has non-vanishing tangle}. 

We also find that $N$-qubit GHZ-like states 
\begin{equation}
\label{Nq}
\vert\Psi_N\rangle= 
\sqrt{p}\, \vert 0,0,\ldots , 0 \rangle+e^{i\, \phi}\, \sqrt{1-p}\,  \vert 1,1,\ldots , 1 \rangle], 
 \end{equation} 
(where $0\leq p\leq 1,\ 0\leq \phi \leq 2\pi$)  violate the separability bound (\ref{vio}) on moments for all 
values of $p\neq 0,1,$  as 
\begin{eqnarray}  
\langle\Sigma_1\rangle&=&2\, \sqrt{p(1-p)}\, \cos\phi, \nonumber \\
\langle\Sigma_2\rangle&=&2\, \sqrt{p(1-p)}\, \sin\phi, \nonumber \\ 
 \langle\Sigma_3^{\rm PT}\rangle&=& 0,\  \ \  
 \langle\Sigma_0^{\rm PT}\rangle=0.    
\end{eqnarray}
in the state (\ref{Nq}) -  with  unit vectors $(k_\alpha,l_{\alpha},m_\alpha)$ chosen as in (\ref{uv}).

We now show that the PPT bound (\ref{vio}) can be used to detect entanglement in mixed states too.   
For this purpose we consider a $N$-qubit Werner state, 
\begin{equation}
\label{rhower}
\hat\rho_{N}(x)=x\, \vert {\rm GHZ}\rangle_N\langle {\rm GHZ}\vert+ \frac{(1-x)}{2^N}\, 
\bigotimes_{\alpha=1}^{N} I ,  
\end{equation}
where 
$\vert {\rm GHZ}\rangle_N=\frac{1}{\sqrt{2}}\, [\vert 0,0,\ldots , 0 \rangle+\vert 1,1,\ldots , 1 \rangle]$  
and  $0~\leq~x~\leq~1.$ 
Fixing the unit vectors as in (\ref{uv}) and performing PT operation on the first qubit, we obtain,  
\begin{eqnarray}
\langle\Sigma_1^{\rm PT}\rangle &=& -x,\ \   \langle \Sigma_2^{\rm PT}\rangle=0,\nonumber \\  
\langle\Sigma_3^{PT}\rangle&=& 0,\  \ 
\langle(\Sigma_0^{\rm PT})\rangle= \frac{(1-x)}{2^{N-1}}. 
\end{eqnarray} 
The PPT bound (\ref{vio}) is satisfied by the multiqubit Werner state if and only if 
\begin{equation}
0\leq x \leq \frac{1}{2^{N-1}+1},
\end{equation}  
which is the necessary and sufficient condition~\cite{PR} for separability of the state (\ref{rhower}). 
It is worth pointing out here that  entanglement witness employed in Ref.~\cite{Toth}   
leads to weaker regimes of inseparability  for this state. Thus, it is significant that a novel set of  
composite multiqubit moments given by Eqs.~(\ref{sigma}),(\ref{sigma3}) capture 
the inseparability behavior of the state (\ref{rhower}) completely.

\section{Summary}
We have analyzed bounds imposed on matrix of multiqubit moments due to positivity 
under partial transpose of density operator. These PPT bounds, in general, place 
additional stronger restrictions on the moments,  
than the conventional ones imposed due to positivity of the quantum state.  
While the set of all separable states obey PPT bounds on moments, violation of these constraints  
are sufficient to detect entanglement in bipartite divisions of the density operator. 
By constructing an appropriate PT matrix of moments 
we have recovered  inseparability conditions of Horodecki et. al.,~\cite{Hor} 
for entangled two qubit states. 
We have also investigated a generalized matrix of moments for multiqubit states 
and derived its PPT bounds. It is shown that an arbitrary pure three qubit state violates  
PPT restrictions on this generalized moment matrix if and only if its three tangle~\cite{Coffman} is non-zero.   
As yet another consequence of violation of these PPT bounds, we recover the necessary and sufficient 
condition for entanglement in $N$ qubit Werner state. It is for the first time that inequalities 
involving  composite multiqubit moments are shown to capture the complete inseparability status of this 
important class of mixed states. 

\section*{ACKNOWLEDGEMENTS} 

We thank Professor G. S. Agarwal for making available a copy of the preliminary write-up of 
their e-print Ref.~[6].  ARU thanks the Commonwelath Commission, UK for supporting this work through the 
award of an Academic Fellowship and also gratefully acknowledges the hospitality extended by the 
University of Bristol.

\end{document}